# An Automatic Identification System (AIS) Database for Maritime Trajectory Prediction and Data Mining

Shangbo Mao[1], Enmei Tu[1,*], Guanghao Zhang[2], Lily Rachmawati[3], Eshan Rajabally[4], and Guang-Bin Huang[2]

[1]Rolls-Royce@NTU Corporate Lab, Nanyang Technological University, Singapore
`{SBMAO, emtu}@ntu.edu.sg`
[2]School of Electrical & Electronic Engineering, Nanyang Technological University, Singapore
`GZHANG009@e.ntu.edu.sg, EGBHuang@ntu.edu.sg`
[3]Computational Engineering Team, Advanced Technology Centre, Rolls-Royce Singapore Pte Ltd
`lily.rachmawati@rolls-royce.com`
[4]Strategic Research Center, Rolls-Royce Plc
`Eshan.Rajabally@Rolls-Royce.com`

**Abstract.** In recent years, maritime safety and efficiency become very important across the world. Automatic Identification System (AIS) tracks vessel movement by onboard transceiver and terrestrial and/or satellite base stations. The data collected by AIS contain broadcast kinematic information and static information. Both of them are useful for maritime anomaly detection and vessel route prediction which are key techniques in maritime intelligence. This paper is devoted to construct a standard AIS database for maritime trajectory learning, prediction and data mining. A path prediction method based on Extreme Learning Machine (ELM) is tested on this AIS database and the testing results show this database can be used as a standardized training resource for different trajectory prediction algorithms and other AIS data based mining applications.

## 1 Introduction

In the modern globalized economy, ocean shipping becomes the most efficient method for transporting commodities over long distance. The persistent growth of the world economy leads to increasing demand of maritime transportation with larger ship capacity and higher sailing speed [1]. In this circumstance, safety and security become key issues in maritime transportation. Intelligent maritime navigation system using Automatic Identification System (AIS) data improves the maritime safety with less cost compared with conventional maritime navigation system using human navi-gators. The AIS is a maritime safety and vessel traffic system imposed by the International Maritime Organization (IMO). Autonomously broadcasted AIS messages contain kinematic information (including ship location, speed, heading, rate of turn, destination and estimated arrival time) and static information (including ship name, ship MMSI ID, ship type, ship size and current time), which can be transformed into useful information for intelligent maritime traffic manipulations, e.g. vessel path prediction and collision avoidance, and thus plays a central role in the future autonomous maritime navigation system. Over the last several years, receiving AIS messages from vessels and coastal stations has become increasingly ordinary.

---

* Corresponding author: Enmei Tu, hellotem@hotmail.com

Although sufficient AIS data can be obtained from many data providers, e.g. Marinecadastre (MarineC.) [3] and Sailwx [4], to the best of our knowledge, there is no existing standard AIS benchmark database in maritime research area, which makes it quite inconvenient for researchers and practitioners in the field, since collecting a usable dataset will cost a lot of time and effort. Furthermore, as the intelligent maritime system develops rapidly, many researchers proposed anomaly detection and motion prediction algorithms and it is quite important to have a database that could be served as a benchmark for comparing the performance of different methods and algorithms. For example, in 2008, B. Ristic et al. [9] proposed an anomaly detection and motion prediction algorithm based on statistical analysis of motion pattern in AIS data. In 2013, Premalatha Sampath generated vessel trajectory from raw AIS data and analyzed the trajectory to identify the kinematic pattern of vessel in New Zealand waterways [10]. So in this paper, a ready-to-use standard AIS database is constructed for maritime path learning, prediction and data mining.

The remaining parts of the paper are organized as follows: Section 2 describes the AIS data type and data source. Section 3 describes the detailed process of constructing the AIS database. Then the structure and static information of our AIS database are summarized and described in Section 4. Finally, we conduct an experiment based Extreme Learning Machine (ELM) on the AIS database to show the usefulness of it in Section 5.

## 2 Properties of AIS Database

This section describes the attributes of AIS data and introduces some popular AIS data providers. And AIS data have some special attributes which would lead to the difference between maritime trajectory prediction and route prediction in other fields.

### 2.1 AIS Data Attributes

AIS technology broadcast ship information and voyage information at regular time interval. The information can be received by onboard transceiver and terrestrial and/or satellite base station. There are some important attributes of AIS data: longitude, latitude, speed over ground (SOG), course over ground (COG), vessel's maritime mobile service identity (MMSI), base date time, vessel type, vessel dimension, rate of turn (ROT), navigation status and heading. In this paper, the standard AIS database contains longitude, latitude, SOG, COG, MMSI and base date time, which are the most useful attributes for maritime trajectory learning and prediction.

### 2.2 AIS Data Providers

There are many existing AIS data providers e.g. Marine Traffic (Marine T) [12], VT explorer (VT E.) [13], FleetMon [16], Marinecadastre (MarineC.) [3] and Aprs [7]. Among these providers, MarineC can be downloaded for free and have good data quality according to data completeness and position precision. So in this paper, MarineC is selected to collect AIS data online. MarineC contains historical records from 2009 to 2014 in America at a minute interval. We can choose and download AIS data files in specific month and specific interest area. We downloaded February 2009 AIS data in UTM zone ten. However, the AIS data downloaded from MarineC contains some data missing. To solve these problem, we use the linear interpolation and the detail will he introduced in this paper later.

## 3   AIS Database Construction

This section describes the data processing tool and the detail of constructing the standard AIS database we proposed. The whole process contains four parts: raw data pre-processing, raw data selecting, candidate data cleaning and missing data interpolating.

### 3.1   Raw Data Pre-Processing

The first step of constructing an AIS database is to download the raw database file with dbf format from http://www.marinecadastre.gov/ais/. Prior to download raw data online, the selection of interest area is necessary. As shown in Fig. 1, zone ten is on west coast of the United States and it contains considerable amount of ships. And these AIS data are open-sourced. In this paper, zone ten is chosen as interested area because it contains sufficient amount of AIS data. The shaded part in Fig. 1 is the chosen interested area whose longitude is from -120 to -126 degree and latitude is from 30 to 50 degree.

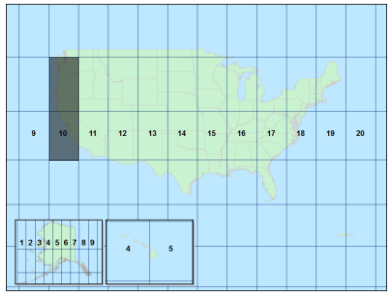

**Fig. 1.** UTM zone map and data source location

In order to pre-process the AIS data and pick out the useful data, an application which can transfer raw database file with dbf format to csv (comma-separated values) format is required since csv format file is constituted by lines of tabular data records and is much easier to be handled by researchers. Arcmap is the most frequent cited Geographic Information System (GIS) software and is mainly used to view, edit, create and analyze geospatial data. Since Arcmap is selected as our data transformation software, a tool of it named "export feature attribute to ASCII" was used for exporting feature class coordinates and attribute values to a space, comma, or semicolon-delimited ASCII text file. The exporting result is presented in Fig. 12 in Section 4.

### 3.2 Raw Data Selecting

After raw data pre-processing, it is necessary to select the candidate data from the raw data with Excel format. Data selection contains two steps. First, for manipulation convenience in the following operations, the whole raw data are sorted in increasing time order and then sorted again by MMSI. One MMSI represents one single vessel. Thus in this way, the track of each single ship can be displayed in chronological order and easier to process. The second step is to calculate route complexity and longest duration of navigation.

- Longest duration of navigation
  If the SOG value of a vessel meets the following inequality, we call this vessel is in the navigation condition.

$$SOG \neq 0 \tag{1}$$

Therefore, the longest duration of navigation is defined as the longest continuous nonzero SOG sequence of the AIS messages in the navigation condition. As a standard AIS database for maritime trajectory prediction and data mining, one single route should contain substantial information. Thus, the route with short duration which contains not enough data for training and testing on the route prediction and data mining algorithms. Based on our experience, the selection requirement of this property is that the trajectory data contain more than 500 AIS messages.

- Route complexity

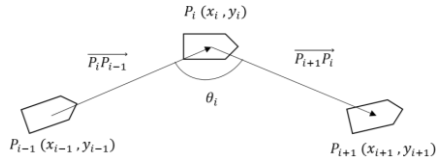

**Fig. 2.** Sample of route complexity

For each single route, the $\cos\theta$ of each ship position is calculated and the definition of route complexity is the mean value of $\cos\theta$, which can be calculated by the following equation:

$$\cos\theta_i = \frac{\overrightarrow{P_iP_{i-1}} \bullet \overrightarrow{P_{i+1}P_i}}{\left|\overrightarrow{P_iP_{i-1}}\right|\left|\overrightarrow{P_{i+1}P_i}\right|} \quad (2)$$

Where $P_i(x_i, y_i)$ is vessel position at $T_i$; $x_i$ is the longitude of vessel at $T_i$; $y_i$ is the latitude of vessel at $T_i$ and $\overrightarrow{P_iP_{i-1}}$ is the vector $(x_i - x_{i-1}, y_i - y_{i-1})$. The route complexity should be larger than 0.8 in our database since the trajectory whose complexity is lower than 0.8 is tangled.

### 3.3 Candidate Data Cleaning

After candidate data were obtained, further selection based on trajectories is required. All trajectories of candidate data are plotted by MATLAB. Among all the trajectories, we defined three noisy trajectory types as follows. (showed from Fig. 3 to Fig. 6 and the horizontal axis is longitude and the vertical axis is latitude), we then removed them all:

- The discontinuous trajectory is showed in Fig. 3 and Fig. 4.
- The loose trajectory is showed in Fig. 5.
- The tangled trajectory is shown in Fig. 6.

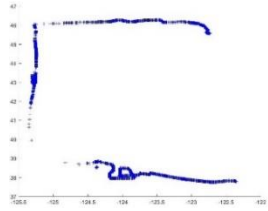  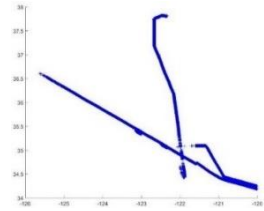

**Fig. 3.** Discontinuous trajectory sample-1     **Fig. 4.** Discontinuous trajectory sample-2

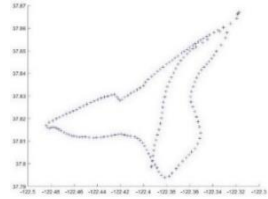  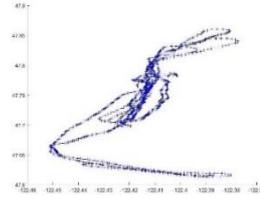

**Fig. 5.** Loose trajectory sample     **Fig. 6.** Tangled trajectory sample

Because these noisy trajectories have some inherent drawbacks. Routes prediction and data mining algorithms cannot learn the patterns of routes. And the shapes of noisy routes are not typical. Once the noisy trajectories are identified, they should be removed. Finally, 200 useful trajectories which contain 403599 AIS records are saved and used to construct the standard AIS database. Fig. 7 shows some typical trajectories in our database (The horizontal axis is longitude and the vertical axis is latitude).

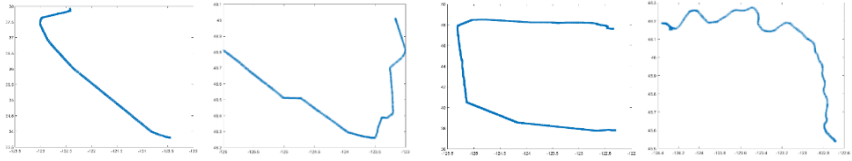

**Fig. 7.** Reserved trajectory sample

### 3.4 Missing Value Interpolating

In our database, the discontinuousness caused data missing values may affect the performance of learning algorithms and data mining quality of the database. Besides, the raw data also contain erroneous speed data. Before performing interpolation, we have to detect and remove the erroneous data. Each AIS record represents position of the ship. There are 403599 ship positions in the database. The detection of speed errors is based on SOG jump (the difference between current SOG and previous SOG). If the jump is larger than the threshold we set in advance, we calculate the distance between the two messages using the latest speed and test if this distance is consistent with the actual distance between the messages given by Haversine formula [18], i.e. the calculated distance should be close to the actual distance within a small threshold if the speed jump is correct. If not so, the latest speed is treated as erroneous and is set to previous speed. The Harversine formula is showed below. $d$ is the distance between two points with longitude and latitude $(\psi, \phi)$ and $r$ is the radius of Earth.

$$d = 2r\sin^{-1}(\sqrt{\sin^2(\frac{\phi_2 - \phi_1}{2}) + \cos(\phi_1)\cos(\phi_2)\sin^2(\frac{\psi_2 - \psi_1}{2})}) \qquad (3)$$

The second row in Fig. 8 is an example of incorrect SOG jump. In order to interpolate the missing values efficiently, all of the AIS records with speed errors should be corrected in advance.

For path interpolation, there are three steps: detecting data missing, judging if it needs interpolation and making linear interpolation. Data missing occurs when the time interval period between two consecutive messages is larger than one chosen interval. We choose one minute as the threshold interval in this paper. Once detected, these two row data are defined as the missing data pair. A sample of missing data pair is shown in Fig. 9 in which there is a five-minute interval between the two consecutive messages. Then the missing time period is defined as the time range between the missing data pair and the great-circle distance between missing data pair calculated by Haversine formula [18]. The computed distance divides the SOG (km/minute) of the earlier position. The division result, that is larger than two, requires linear interpolation [26]. Fig. 10 and Fig. 11 show a trajectory example before and after the interpolation (From Fig. 10 to Fig. 11, the horizontal axis is longitude and the vertical axis is latitude).

| XCoord | YCoord | SOG | COG | time | MMSI |
|---|---|---|---|---|---|
| -121.1481 | 34.825067 | 20 | 330 | 200901071138 | 366882000 |
| -121.151967 | 34.830567 | 102 | 360 | 200901071139 | 366882000 |
| -121.155453 | 34.83544 | 21 | 329 | 200901071140 | 366882000 |

**Fig. 8.** Example of erroneous SOG jump

| XCoord | YCoord | SOG | COG | ROT | BASEDATETIME | MMSI |
|---|---|---|---|---|---|---|
| -124.9991 | 43.2833 | 12 | 359 | 0 | 200902011307 | 258919000 |
| -124.999217 | 43.298783 | 12 | 0 | 0 | 200902011311 | 258919000 |

**Fig. 9.** Exmaple of missing data pair

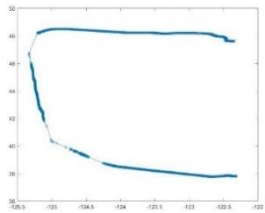
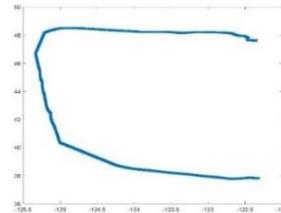

**Fig. 10.** Before interpolation  **Fig. 11.** After interpolation

The principle of linear interpolation is that we presume that the ship is in uniform linear motion during the missing time period and the speed is considered as the SOG of the earlier position. The calculation and interpolation of the missing data are based on these two assumptions.

## 4    Description of AIS Database

In this section, we introduce the standard AIS database we constructed in two parts: the structure and statistical information of the database.

### 4.1    Structure

The whole AIS database contains 200 clean trajectories stored in 200 csv file[5]. Each file is named by the MMSI and sorted in increasing time order. Each csv file contains latitude, longitude, SOG, COG, ROT, time and MMSI of a single ship. Fig. 12 presents part of one xlsx file as an example.

| XCoord | YCoord | SOG | COG | ROT | BASEDATETIME | MMSI |
|---|---|---|---|---|---|---|
| -120.003717 | 34.242683 | 20 | 285 | 0 | 200902012013 | 235844000 |
| -120.0102 | 34.244133 | 20 | 285 | 0 | 200902012014 | 235844000 |
| -120.016783 | 34.245633 | 20 | 285 | 0 | 200902012015 | 235844000 |

**Fig. 12.** Example of a csv file

---

[5]  The files will be uploaded to UCI Machine Learning repository (http://archive.ics.uci.edu/ml/).

## 4.2 Statistical Information

The raw data was chosen from limited area and time periods. The range of longitude is from -120 to -126 degree and the latitude is from 30 to 50 degree as we showed in Fig. 1. The complexities of routes are larger than 0.87.

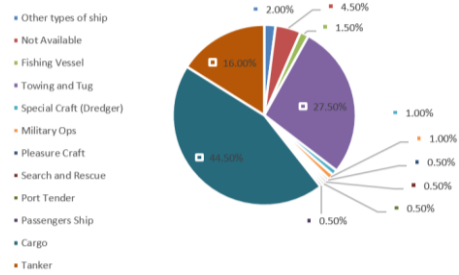

**Fig. 13.** Distribution of vessel types

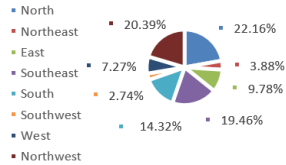

**Fig. 14.** Distribution of COG

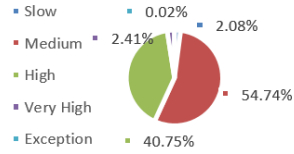

**Fig. 15.** Distribution of SOG

**Table 1.** List of COG statuses

| Course over ground (COG) | Statuses |
|---|---|
| [337.5, 360] ∪ [0, 22.5) | North |
| [22.5, 67.5) | Northeast |
| [67.5, 112.5) | East |
| [112.5, 157.5) | Southeast |
| [157.5, 202.5) | South |
| [202.5, 247.5) | Southwest |
| [247.5, 292.5) | West |
| [292.5, 337.5) | Northwest |

**Table 2.** List of SOG statuses

| Speed over ground (SOG) | Statuses |
|---|---|
| [0, 3) | Slow |
| [3, 14) | Medium |
| [14, 23) | High |
| [23, 99) | Very high |
| Over 99 | Exception |

There are many ship types within this database including cargo ships, tankers, tugs, ships engaged in military operations, etc. The detail distribution of vessel types is shown in the Fig. 13. COG is the actual direction of a vessel and is often affected by the weather over sea. All AIS data are divided into eight statuses according to the COG values as we showed in Table 1 [17]. Fig. 14 shows the distribution of vessels' direction. SOG represents the speed of vessel and is an important parameter for us to analyze the AIS data. Five statues of SOG are defined [17] according to AIS dynamic information and listed in Table 2. The percentage of each SOG status is showed in

Fig. 15. The length of each route is another important property of AIS data and could help us analyze the route length. According to the data amount of each ship, all two hundred ship routes are separated into four types: short route, medium route, long route and exception route as we listed in Table 3. Fig. 16 shows the proportion of each route type with different trajectory length. After linear interpolation, the length of each trajectory has changed so we summarize the distribution of interpolated trajectories in Fig. 17, from which we can see that most of the trajectories in the processed database belong to medium and long categories. Fig. 18 is the histogram of how many positions are interpolated. (The horizontal axis is interpolated length and the vertical axis is number of trajectories for each interpolated length).

**Table 3.** List of route types

| Data quantity range | Route types |
|---|---|
| [530, 1000) | Short |
| [1000, 2000) | Medium |
| [2000, 10000) | Long |
| Over 10000 | Exception |

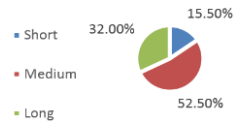
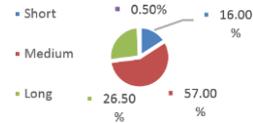

**Fig. 16**. Length of original route distribution       **Fig. 17.** Length of interpolated route distribution

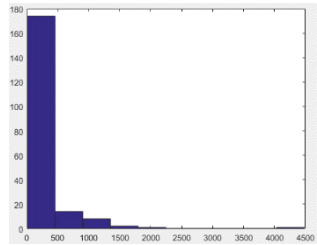

**Fig. 18.** Histogram of interpolated length

## 5    Experiments

In this section, we give an experimental demonstration of the usefulness of this database and summarize the experimental results. The experiments are conducted for maritime trajectory learning and prediction that are aimed to predict future motion of ship based on the current position and historic movement. Extreme Learning Machine (ELM) is a learning algorithm for single hidden layer feedforward networks (SLFNs) with random hidden nodes [20] [21]. ELM have good generalization performance and fast learning speed. As a preliminary experiment, we run ELM on our AIS database to predict the vessel trajectory. In the future, we will compare more different algorithms on our AIS database e.g. Kalman Filter, Gaussian Process, etc.

AIS data are time series data and new data of each feature comes continuously with potentially uneven time interval. It is a question how these machine learning algorithms could make the utmost use of historical data to predict different future vessel position with a dynamic prediction time. In this project, we use the following data segmentation method, as illustrated in Fig. 19. Suppose the training set contains $s$ samples and each sample feature has length $l$. The prediction time is $t_p$. We start at current time $t_c$, indicated by red line. The first training sample is cut at time tick $t_c$-$t_p$-$l$ with time length $l$ and its target value is vessel position at time $t_c$. The second training sample is cut at time tick $t_c$-$t_p$-$l$-1 with time length $l$ and its target value is vessel position at time $t_c$-1 and so on for the rest of all training samples. The testing sample is cut backwards at $t_c$ with feature length $l$. This method makes sure that the latest data can be utilized for both training and testing, without any dependence to future information.

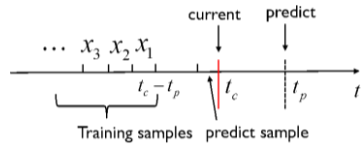

**Fig. 19.** Sample segmentation of trajectory prediction

ELM is used to make the trajectory prediction in these experiments. We analyze the performance of the algorithm according to these testing results. In order to make a comprehensive evaluation of ELM, predicting the same trajectory in 20 minutes and 40 minutes is performed in this experiment.

The testing results contain two parts: prediction results and error distribution which are described and discussed accordingly in this section. The prediction results show the original and predicted trajectory and algorithm performance in direct way. The prediction results of 20 minutes and 40 minutes interval are presented in the Fig. 20 and Fig. 21. From these two figures, we can find the ELM performance of 20 minutes experiment is much better than 40 minutes one. Since the vessel motion is often affected by the dynamic and unpredictable sea weather, the task of predicting the route in 40 minutes is more challenging and complex. (From Fig. 20 to Fig. 21, the horizontal axis is longitude and the vertical axis is latitude)

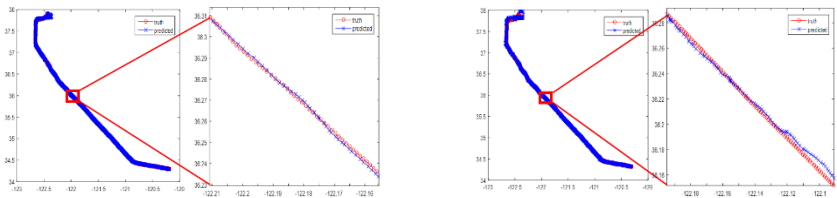

**Fig. 20.** Prediction results: 20 min      **Fig. 21**. Prediction results: 40 min

The error distribution helps us to analyze the algorithm performance in a quantitative way. Error is defined as the earth surface distance between real position and predicted position. And this distance can be calculated by Harversine formula which has been showed in Section 3.4. As we showed in Fig. 22, the error is from 0 to 2.5 when the experiment is to predict the trajectory in 20 minutes. Most of them are from 0 to 0.5. Fig. 23 shows the error of 40 minutes distributing from 0 to 6 and most of them are in the range between 0 and 1 (From Fig. 22 to Fig. 23, the horizontal axis is average error (Nautical mile) and the vertical axis is number of predictions for each error).

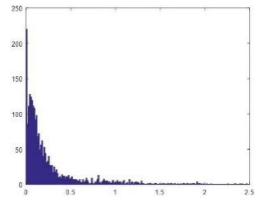 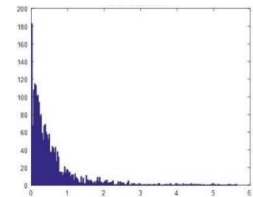

**Fig. 22.** Error distribution: 20-minute prediction results      **Fig. 23**. Error distribution: 40-minute prediction results

## 6 Conclusions

In conclusion, in this paper we present a standard AIS database and described the details of constructing it. The testing results on our AIS database demonstrate its potential value of serving as a benchmark for maritime trajectory learning, prediction and data mining. Our following work will be focused on conducting more experiments on the database by including more machine learning algorithms, such as manifold clustering algorithm [22] and semi-supervised learning algorithms [23]. In the future, this database can also be used as a benchmark database to verify the efficiency of other novel AIS data mining algorithms and compare their performances.

## 7 References

1. P. Kaluza, A. Kölzsch, M. T. Gastner, and B. Blasius.: The complex network of global cargo ship movements. In: Journal of the Royal Society Interface, vol. 7, no. 48, pp. 1093–1103 (2010)
2. G. Pallotta, M. Vespe, K. Bryan.: Vessel Pattern Knowledge Discovery from AIS Data: a Framework for Anomaly Detection and Route Prediction. In: Entropy, vol. 6, issue 15, pp. 2218-2245 (2013)
3. The Marinecadastre website. [Online]. Available: `http://marinecadastre.gov/ais/`
4. The Sailwx website. [Online]. Available: `http://sailwx.info/`
5. The Aishub website. [Online]. Available: `http://aishub.net`

6. The Ais Exploratorium website. [Online]. Available: http://ais.exploratorium.edu/
7. The Aprs website. [Online]. Available: http://aprs.fi/page/api
8. W. M. Wijaya and Y. Nakamura.: Predicting Ship Behavior Navigating through Heavily Trafficked Fairways by Analyzing AIS Data on Apache HBase. In: 2013 First International Symposium on Computing and Networking (CANDAR), pp. 220–226 (2013)
9. B. Ristic, B. La Scala, M. Morelande, and N. Gordon.: Statistical Analysis of Motion Patterns in AIS Data: Anomaly Detection and Motion Prediction. In: 11th International Conference on Information Fusion, pp.1–7 (2008)
10. Premalatha Sampath and David Parry.: Trajectory Analysis using Automatic Identification System in New Zealand Waters. In: International Journal of Computer and Information Technology, vol. 2, pp. 132-136 (2013)
11. The NAVIGATION CENTER website. [Online]. Available: http://www.navcen.uscg.gov/?pageName=AISmain
12. The Marine Traffic website. [Online]. Available: http://www.marinetraffic.com/
13. The VT Explorer website. [Online]. Available: http://www.vtexplorer.com/
14. The IHS Global website. [Online]. Available: https://www.ihs.com/index.html
15. The exactEarth website. [Online]. Available: http://www.exactearth.com/
16. The FleetMon website. [Online]. Available: https://www.fleetmon.com/
17. F. Deng, S. Guo, Y. Deng, H. Chu, Q. Zhu and F. Sun.: Vessel track information mining using AIS data. In: Multisensor Fusion and Information Integration for Intelligent Systems (MFI), 2014 International Conference on. IEEE, pp. 1-6 (2014)
18. Chopde, Nitin R., and M. Nichat.: Landmark based shortest path detection by using A* and Haversine formula. In: International Journal of Innovative Research in Computer and Communication Engineering, vol. 1, no. 2, pp. 298-302 (2013)
19. Eriksen, Torkild, et al.: Maritime traffic monitoring using a space-based AIS receiver. In: Acta Astronautica, vol. 58, issue 10, pp. 537-549 (2006)
20. G. B. Huang, Q. Y. Zhu, and C. K. Siew.: Extreme learning machine: Theory and applications. In: Neurocomputing, vol. 70, no. 1–3, pp. 489–501 (2006)
21. G.-B. Huang, H. Zhou, X. Ding, and R. Zhang.: Extreme learning machine for regression and multiclass classification. In: IEEE Transactions on Systems, Man, and Cybernetics, Part B (Cybernetics), vol. 42, no. 2, pp. 513–529 (2012)
22. Tu, Enmei, Longbing Cao, Jie Yang, and Nicola Kasabov.: A novel graph-based k-means for nonlinear manifold clustering and representative selection. In: Neurocomputing, vol. 143 pp. 109-122 (2014)
23. Tu, Enmei, Yaqian Zhang, Lin Zhu, Jie Yang, and Nikola Kasabov.: A Graph-Based Semi-Supervised k Nearest-Neighbor Method for Nonlinear Manifold Distributed Data Classification. arXiv preprint arXiv: 1606.00985 (2016)
24. Jackson, Michael R., Yiyuan J. Zhao, and Rhonda A. Slattery.: Sensitivity of trajectory prediction in air traffic management. In: Journal of Guidance, Control, and Dynamics, vol. 22, no. 2, pp. 219-228 (1999)
25. Liu, Tong, Paramvir Bahl, and Imrich Chlamtac.: Mobility modeling, location tracking, and trajectory prediction in wireless ATM networks. In: IEEE Journal on selected areas in communications, vol. 16, no. 6, pp. 922-936 (1998)
26. Meijering, Erik.: A chronology of interpolation: from ancient astronomy to modern signal and image processing. In: Proceedings of the IEEE, vol. 90, no. 3, pp.319-342 (2002).